# First order antiferro-ferromagnetic transition in Fe$_{49}$(Rh$_{0.93}$Pd$_{0.07}$)$_{51}$ under simultaneous application of magnetic field and external pressure

Pallavi Kushwaha, Pallab Bag, R Rawat*, and P Chaddah

UGC-DAE Consortium for Scientific Research, University Campus, Khandwa Road, Indore-452001, India.

The magnetic field-pressure-temperature (H-P-T) phase diagram for first order antiferromagnetic (AFM) to ferromagnetic (FM) transition in Fe$_{49}$(Rh$_{0.93}$Pd$_{0.07}$)$_{51}$ has been constructed using resistivity measurements under simultaneous application of magnetic field (up to 8 Tesla) and pressure (up to 20 kbar). Temperature dependence of resistivity (ρ-T) shows that with increasing pressure, the width of the transition and the extent of hysteresis decreases whereas with the application of magnetic field it increases. Consistent with existing literature the first order transition temperature (T$_N$) increases with the application of external pressure (~ 7.3 K/ kbar) and decreases with magnetic field (~ -12.8 K/Tesla). Exploiting these opposing trends, resistivity under simultaneous application of magnetic field and pressure is used to distinguish the relative effect of temperature, magnetic field and pressure on disorder broadened first order transition. For this a set of H and P values are chosen for which T$_N$ (H$_1$, P$_1$) = T$_N$ (H$_2$, P$_2$). Measurements for such combinations of H and P show that the temperature dependence of resistivity is similar i.e. the broadening (in temperature) of transition as well as extent of hysteresis remains independent of H and P. The transition width decreases exponentially with increasing temperature. Isothermal magnetoresistance measurement under various constant pressure show that even though the critical field required for AFM-FM transition depends on applied pressure, the hysteresis as well as transition width (in magnetic field) both remains independent of pressure, consistent with our conclusions drawn from ρ-T measurements.

___________________

*Email: rrawat@csr.res.in

# I. INTRODUCTION

The first order antiferromagnetic (AFM) to ferromagnetic (FM) transition in FeRh systems have been subject matter of intensive theoretical and experimental investigations [1–4]. This transition is known to be accompanied with large change in resistivity, volume, entropy, which give rise to giant magnetoresistance [4–7], magnetostriction [8–10] and magnetocaloric effect [11–13]. Since there is a large change in volume associated with the first order magnetic transition (~ 1%) [9], it can be tuned or induced with the application of external pressure [14, 16, 17]. The transition temperature ($T_N$) is shifted to higher temperature with the application of external pressure as the low temperature AFM phase has lower volume compared to high temperature FM phase e.g. in case of $Fe_{47.7}(Rh_{0.97}Pd_{0.03})_{52.3}$ the rate of change of AFM-FM transition temperature ($T_N$) is ~ 6.55 K/kbar, whereas the FM-paramagnetic (PM) transition temperature ($T_C$) shifts to lower temperature at the rate of -0.90 K/kbar [14]. Wayne [14] pointed out a correlation between $T_N$ and $dT_N/dP$ in that the lower the $T_N$ larger is the $dT_N/dP$. By comparing his high pressure data with magnetic field dependence of $T_N$ by Kouvel et al. [18], Wayne found that the ratio of $dT_N/dH$ and $dT_N/dP$ is almost constant (~ −1.75 kbar/Tesla). Similar increase of transition temperature with applied pressure was observed for $(Fe_{1-x}Ni_x)_{49}Rh_{51}$ [15–17]. Since the $T_N$ is lower for these compounds they also have larger $dT_N/dP$ values consistent with earlier studies. These studies address the variation of transition temperature with pressure or magnetic field but other characteristic of first order transition e.g. hysteresis width as well as transition width remained unaddressed.

Hysteresis and transition width are one of the characteristics of a first order transition and has implication for practical application of such materials. The quench disorder in substitutional alloys and compounds leads to broadening or smearing of first order magnetic transition [19] and result in phase coexistence around the transition region [20]. Such disorder broadened transitions are of interest in a variety of systems e.g. intermetallics [4, 21, 22], oxides [23–27], shape memory alloy [28–31], etc. Beside quench disorder magnetic field, pressure and strain are known to influence broadening or smearing of transition. In recently discovered glass like magnetic states [32], the disorder broadening of first order transition lead to tunable phase fraction of glass like magnetic state at low temperature [33]. The broadening or smearing of transition with composition and external parameter is also pursued in the study of quantum phase transitions [34]. However there are very few studies which addresses the relative role of external parameter in the broadening of first order phase transition.

Here we report our study of first order transition in Pd doped FeRh under simultaneous application of pressure and magnetic field. We have used opposing influence of pressure and magnetic field on first order transition to distinguish relative role these parameters and temperature in determining the broadening of transition and the extent of hysteresis.

# II. EXPERIMENTAL DETAILS

Polycrystalline sample used in the present study is the same as that used in the previous study [4], except for one additional annealing at $950^0$ C for 12 hours in vacuum. This annealing has resulted slight increase in the peak associated with chemically disordered phase, though the first order transition temperature ($T_N$ ~ 212K) remains almost same as that reported for our earlier sample [4]. The resistivity measurements under applied external pressure (up to 20 kbar) were carried out using BeCu High pressure cell from M/s. easyLab U.K. along with Oxford superconducting magnet system. Iso amyl alcohol and n pantene in 1:1 volume ratio are used as a pressure transmitting medium. Resistance of manganin wire (fixed near sample) is used for measuring pressure inside the pressure cell. The infield (up to 8 Tesla) measurements were carried out in longitudinal geometry i.e. applied magnetic field direction was kept parallel to the current direction. Resistance as a function of temperature at constant magnetic field and pressure was measured in step mode where resistivity is measured after 5 minutes of temperature stabilization. Care is taken to avoid temperature overshoot during temperature sweep.

## III. RESULTS AND DISCUSSION

Resistivity as a function of temperature is measured under various constant pressure and magnetic field. Figure 1(a) and (b) shows some of these resistivity curves in the presence of various applied pressure for 0 and 8 Tesla magnetic field, respectively. For these measurements pressure and magnetic field are applied at room temperature and measurement is carried out during cooling (FCC) and subsequent warming (FCW). The hysteretic resistivity behavior, which is taken as signature of first order transition, shows that in the absence of applied magnetic field transition is shifted to higher temperature with pressure and at 19.9 kbar no transition is observed i.e. system remains in antiferromagnetic (higher resistivity) state up to room temperature at 19.9 kbar and 0 Tesla. The increase of transition temperature with pressure is expected as the volume of antiferromagnetic phase is smaller (~ 1%) compared to ferromagnetic state around transition temperature [9]. For applied pressure higher than 9 kbar only partial transition from AFM to FM state is observed during warming and the FM phase fraction at 300 K decreases further with increasing pressure. This is shown as inset in figure 1 (a), which shows variation of resistivity (taken from ρ-T data) at 5 K and 300 K. At 300 K sharp change in resistivity is observed between 10 to 14 kbar indicating an increase in AFM phase fraction with pressure and above 14 kbar system appears to be in almost AFM state. The zero field resistivity at 5 K as a function of pressure shows small but distinct increase with pressure above 1.8 kbar. The rise could be due to pressure effect on band structure of this application of magnetic field the resistivity shows a small increase in the AFM state at 5K. The other possibility for this increase in resistivity could be the presence of very small ferromagnetic fraction existing down to lowest temperature and this fraction becomes smaller with increasing pressure. There are reports, both experimental and theoretical, which suggest that the surface/interface (few tens of nanometer) do not show transition and remains ferromagnetic down to lowest temperature [2, 35]. It is possible that the thickness of the untransformed surface/interface decreases with increasing

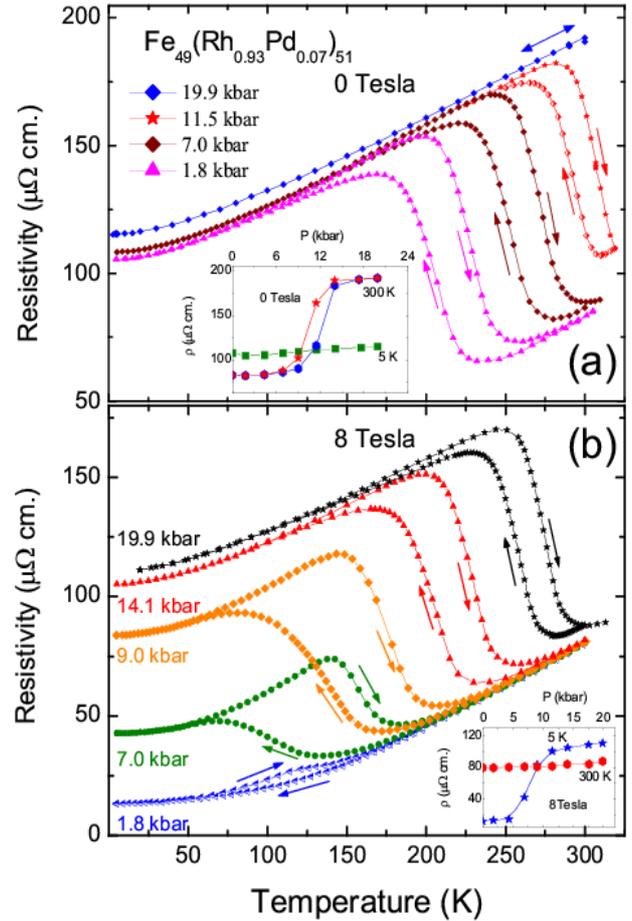

**Figure 1:** Resistivity (ρ) as a function of temperature (T) for $Fe_{49}(Rh_{0.93}Pd_{0.07})_{51}$ at various constant pressure **[a]** in the absence of applied magnetic field and **[b]** in the presence of 8 Tesla magnetic field. Both pressure and magnetic field are applied at room temperature. Inset shows resistivity at 5 K and 300 K obtained from these ρ-T curves as a function of pressure.

pressure resulting in higher AFM phase fraction and therefore higher resistivity.

For 8 Tesla applied magnetic field, almost no transition was observed at low pressure and resistivity value indicates that system is in almost FM state at 5 K, figure 1(b). This is consistent with our earlier report [4] where we do not observe any transition in 8 Tesla magnetic field. With increasing external pressure resistivity shows hysteretic temperature dependence for pressure ≥1.8 kbar. The transition temperature is found to be around 268 K for maximum applied external pressure of 19.9 kbar at 8 Tesla. The resistivity change across the transition remains smaller for pressure ≤11.5 kbar. It indicates that for P ≤11.5 kbar both AFM and FM phase coexist at

5K and FM phase fraction decreases with increasing pressure. The change in resistivity with pressure at 5 K at 8 Tesla magnetic field is shown as inset in figure 1 (b). It indicates an increase in AFM phase fraction with increasing pressure and above 11.5 kbar system is almost in AFM state. The appearance of first order transition with pressure under 8 Tesla magnetic field in our system is similar to that observed in $(Fe_{1-x}Ni_x)_{49}Rh_{51}$ for x= 0.04 to 0.05 composition [16] which do not show transition in the absence of applied external pressure and magnetic field.

A phase diagram has been constructed from resistivity and isothermal magnetoresistance (MR) measurements (to be discussed in the following section), which is shown in figure 2(a). The transition temperature ($T_N$) is taken as the average of the temperature of minimum of $d\rho/dT$ during cooling (T*) and warming (T**). Similarly the critical field required for AFM-FM transition are taken as the magnetic field at which $d\rho/dH$ shows a minima in the isothermal magnetoresistance curve. At low temperature magnetic field required for AFM to FM transition exceeds 8 Tesla. However cooling under 8 Tesla at low pressure give rise to FM state down to 5 K. The isothermal MR measurement for such field cooled condition gives us an estimate of lower critical field (H required for FM to AFM transition). The non-monotonic variation of lower critical field is consistent with our earlier studies in this system [4]. The first order transition temperature observed around 212 K at normal pressure and zero field increases with applied external pressure. Around 212 K the rate of change of transition temperature with pressure and magnetic field is found to be ~7.3 K/kbar and -12.8 K/Tesla, respectively. Therefore ratio of $(dT_N/dH)/(dT_N/dP)$ for our system is -1.75 kbar/Tesla, which is in good agreement with the observation of Wayne [14] for various FeRh systems. Wayne noticed a correlation between $T_N(0)$ and $dT_N/dP$ in FeRh systems; that is lower the $T_N(0)$ higher the $dT_N/dP$. Such correlation of $(dT_N/dP)$ and $(dT_N/dH)$ with $T_N$ is shown Figure 2(b) and (c) respectively. For $T_N \geq 200K$ the rate of change of transition temperature with $T_N$ appears to be linear and our sample is located at the lower end of this linear behavior. For lower $T_N$ the magnitude of $dT_N/dP$ or $dT_N/dH$ shows distinct deviation from this linearity. In case of Ni doped FeRh system [16, 17], where $T_N$ (in the absence of applied magnetic field and pressure) is lower than our system the rate of change of $dT_N/dP$ was found to be higher and deviates from linear trend observed above 200 K. Same appear to be case with Kouvel's data in figure 2(c). In our earlier study [4] we have shown that as the transition is shifted to lower temperature with the application of magnetic field the kinetics of the transition dominates resulting in nonmonotonic variation of lower critical field. Therefore the rise in $dT_N/dP$ and $dT_N/dH$ appears to be consistent with these results. It also suggests that correlation between $T_N$ and $dT_N/dP$ or $dT_N/dH$ is valid irrespective of the fact that $T_N$ is tuned by composition, magnetic field or pressure. A cross section of phase diagram in the P-T space is shown in figure 2 (d) for 0 and 8 Tesla magnetic field. The $T_N$ varies linearly in the absence of magnetic field but shows a non-linear behavior in the presence of 8 Tesla magnetic field. Non linearity in 8 Tesla data is dominant below 200 K as discussed above. The phase diagram in P-H space (figure 2(e)) shows linear variation of critical field with pressure at various constant temperature. However, this slope appears to be smaller and difference between upper and lower critical field appears to be larger for lower T curves.

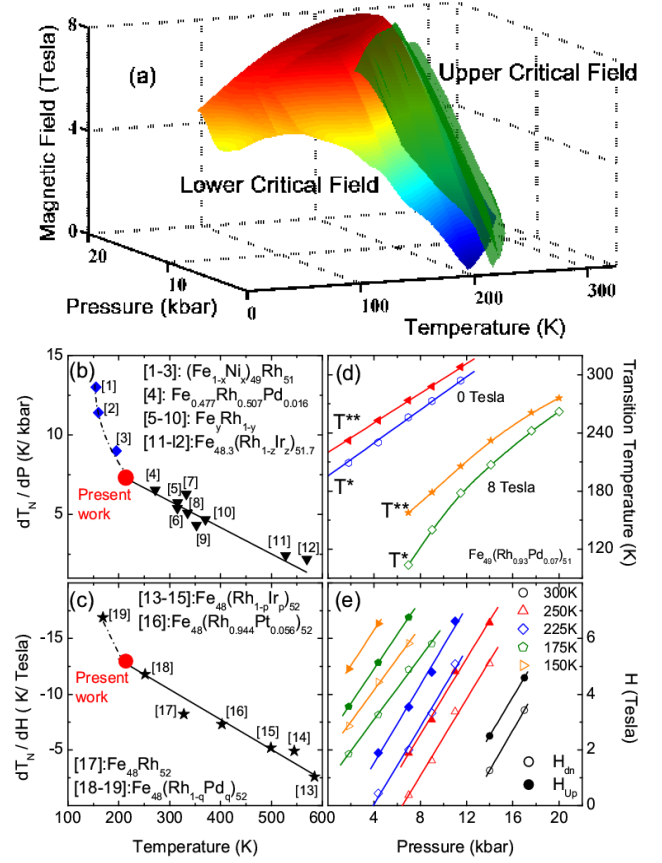

**Figure 2:** *[a] Phase diagram derived from resistivity measurement in temperature, magnetic field and pressure space for $Fe_{49}(Rh_{0.93}Pd_{0.07})_{51}$. [b] $dT_N/dP$ and [c] $dT_N/dH$ as a function of $T_N$ for our sample (red circle) and various FeRh system where x= 0.023 (1), 0.0196 (2), 0.012 (3); y= 0.472 (5), 0.490 (6), 0.48 (7), 0.47 (8), 0.50 (9), 0.50 (10), z= 0.007 (11), 0.112 (12), p= 0.121 (13) 0.084 (14), 0.056 (15); q= 0.029(18), 0.058(19) taken from references [15] for 1-3, [14] for 4-12 and [18] for 13-19. [d] T* and T** vs. P for 0 and 8 Tesla magnetic field [e] Variation of critical field with pressure at various constant temperature.*

From the phase diagram it is also obvious that the hysteresis width in temperature ($H_w \approx$ difference between $T^{**}$ and $T^*$) decreases with pressure and increases with magnetic field, see figure 2 (c) and 2 (d). The increase in $H_w$ with magnetic field, also observed in our earlier work [4], is in qualitative agreement with Baranov et al [5]. Chaddah et al. [36] have investigated the broadening of hysteresis width with external parameter. They concluded that the application of pressure (magnetic field) will lead to broadening of transition if the low temperature phase has lower density (magnetization) compared to high temperature phase and visa versa. This is consistent with our present system where we could verify above prediction for both sign of influence parameter; one for magnetic field and other for pressure. Here, low temperature phase (AFM) has higher density and lower magnetization compared to high temperature phase (FM). The decrease in hysteresis width with increasing magnetic field for paramagnetic-ferromagnetic transition in $La_{0.75}Ca_{0.25}MnO_3$ has been observed by Belevtsev et al. [37] and has been explained within the frame work proposed by Chaddah et al. [36]. They made an empirical observation that this frame work will also be valid for manganites e.g. for AFM-FM transition in $Nd_{0.5}Sr_{0.5}MnO_3$. Recently, Dash et al. [38] studied $La_{0.5}Ca_{0.5}MnO_3$ under simultaneous application of magnetic field and pressure and has shown that pressure asymmetrically affects the first order AFM-FM transition. Our resistivity data shows that not only hysteresis width (difference between $T^*$ and $T^{**}$) but also the transition width during cooling as well as warming depends on external parameter. And resistivity during cooling is more asymmetric compared to during warming at low temperature around $T_N$. To investigate the role of external parameter on hysteresis and transition width, we have used the opposite slope of $dT_N/dP$ and $dT_N/dH$ to obtained same $T_N$ for different set of magnetic field and pressure.

To show the influence of pressure and magnetic field and temperature on transition width we have selected three set of measurements; (i) $T_N \sim$ 260 K, (ii) $T_N \sim$ 200 K and (iii) $T_N \sim$ 150K. Though the applied pressure and magnetic field values are distinctly different, the resistivity behavior is almost identical for all the curves in the set, see figure 3. This is shown more clearly in the temperature derivative of resistivity where all the derivatives in a set overlap with each other. The hysteresis width (difference between the minimum in dρ/dT during cooling and warming) is found to be ~16 K, ~25 K and ~57 K respectively for these three set of measurements. Not only hysteresis width, transition widths are also same for respective cooling as well as warming curve. These results suggest that both pressure and magnetic field can shift the $T_N$ by lowering the free energy of the antiferromagnetic and ferromagnetic state respectively. However, the extent of hysteresis and transition width are determined by the $T_N$.

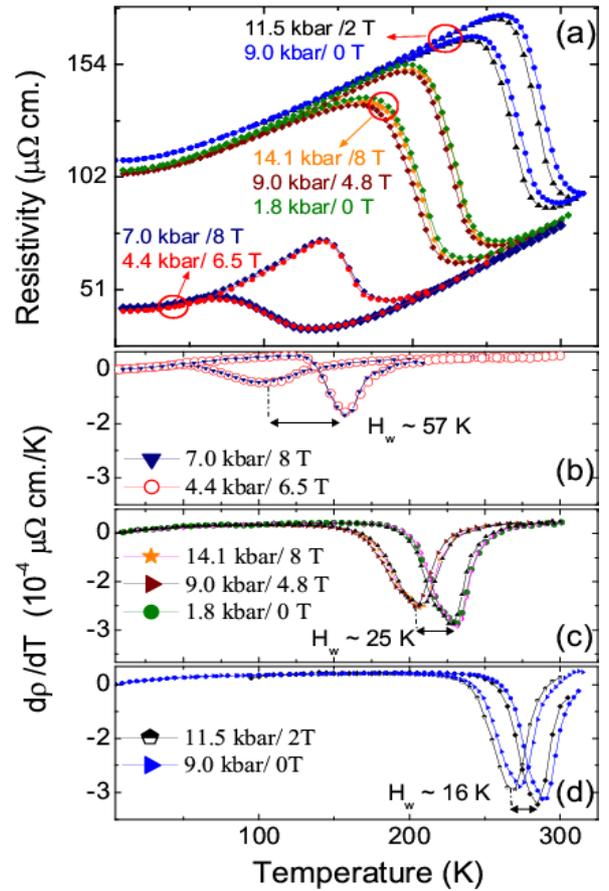

*Figure 3: [a]* Resistivity (ρ) vs. temperature and *[b-d]* dρ/dT vs. T for $Fe_{49}(Rh_{0.93}Pd_{0.07})_{51}$ for three set of P and H values for which $T_N (H_1,P_1) \sim T_N (H_2,P_2)$. These data shows that all the curves in a set overlap with each other even though the applied P and H are different for each curve.

It has been shown that quench disorder leads to spread of local transition temperature [19] and as a result ($H_N$, $T_N$) line for a disorder free sample will be broadened in to a band [39]. Such broadening of the transition in the present system is expected as the transition temperature is known to depend sensitively on the composition as well as chemical ordering of Fe and Rh ions. Each line in the ($H_N$, $T_N$) band correspond to a region of disordered sample with length scale of the order of correlation length [39, 40]. Though the disorder can be considered frozen in our temperature range of measurement, the correlation length decreases with lowering $T_N$. This can result in broadening of ($H_N$, $T_N$) band with lowering temperature. Lower the $T_N$ larger is the broadening. This is shown schematically in figure 4(a), where with increasing magnetic field ($H_{N2} > H_{N1}$) the transition temperature decreases ($T_{N2} < T_{N1}$) and the ($H_N$, $T_N$) band becomes broader. This broadening in temperature (at constant H) is more prominent to the left side of band for $H_{N2} > H_{N1}$. Our dρ/dT data at high temperature suggest gaussian distribution of $T_N$ around mean value, see figure 3(d). This is qualitatively similar to that of Maat et al. [41], where they used two gaussian to fit their temperature and magnetic field derivative of magnetization curve of FeRh films. With these observation, we have used following empirical relation to fit our dρ/dT data

$$\frac{d\rho}{dT} = y_0 - \frac{A}{W}\exp(\frac{-2(T-T_c)^2}{W^2})$$

where,

$$W = w_0(1 + c*\exp(-\frac{T}{t})); \quad A = a(1 + b*T^n)$$

$y_0$ is the slope of resistivity in the AFM or FM state which is taken constant for simplicity, $T_c$ is the temperature where dρ/dT shows a minima (~ T* or T** for cooling and warming curve respectively). The W is the full width at half maxima and empirically assumed to decay exponentially with temperature. The $w_0$, c, t, a, b and n are taken as fitting constant. While fitting the resistivity curves, the parameter W and A are optimized with the constraint that this dependence is true for all the curves. Using this equation we have fitted both heating and cooling dρ/dT curve under various pressure and magnetic field simultaneously. To avoid complication associated with partial transformation at low T, curves which show only partial transformation in our range of T, H and P are excluded from fitting. Some of these fitted curve along with respective experimental dρ/dT curve are shown in figure 4(b). The fitting of these curve gives $w_0$ = 7.77 K, t = 129.81, n= 0.76. Figure 4(c) shows the temperature dependence of transition width W obtained from fitted curve and hysteresis width (T**-T*) taken from phase diagram (figure 2). As can be seen from this figure the transition width in the limit of high temperature will be $w_0 \sim 7.77$ K, which can be considered as the width arising from the quench disorder in the present system. Interestingly, both hysteresis and transition width show a linear variation with $1/T$ (see inset of figure 4 (c)). Using the fitted parameters we have generated resistivity curve for pressure above 11.5 kbar and in the absence of magnetic field, where first order transition temperature was shifted above our temperature range of measurement ($\geq 300$ K). Such simulated curves are shown in figure 4(d). For these curves $T^*$ and $T^{**}$ are taken as linearly extrapolated values of 0 Tesla curve shown in figure 2 (d).

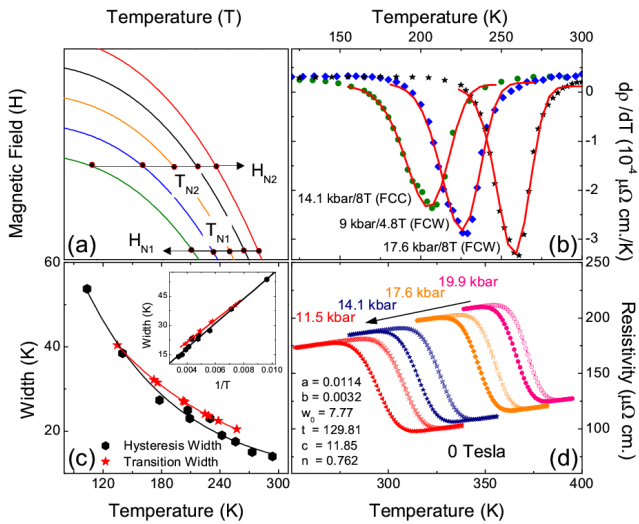

*Figure 4: [a] Schematic diagram showing asymmetric broadening of $T_N$ band at low temperature in H-T space. [b] dρ/dT (symbol) along with fitted curve (red line). [c] Temperature dependence of the transition width W obtained after fitting various resistivity curves along with measured hysteresis width obtained from phase diagram. [d] Simulated temperature dependence of resistivity using equation (see text) at various constant pressure in the absence of applied magnetic field.*

At high temperature the transition width appears to be same for corresponding cooling and warming curve. However, at low temperature we see a distinctly different transition width during cooling and warming. The transition during cooling appears to be much broader compared to that observed during warming. Similar behavior has been observed in many other systems i.e. sharp transition during warming compared to that observed during cooling [21, 22, 28, 38]. To study, if this difference is due to lower $T^*$ compared to $T^{**}$, we have chosen a set of H and P values such that $T^*(H_1, P_1) = T^{**}(H_2, P_2)$. Such three sets of resistivity measurement are shown in figure 5(a) along with their respective temperature derivative (b-d). The increase in transition width with lowering temperature is consistent with above discussion. The temperature derivative of all the curves in a set overlap with each other irrespective of the fact that whether it is taken during cooling or during warming. The overlap of these cooling and warming curve indicate similar nucleation and growth mechanism for transition during cooling and warming. As stated above we have fitted both cooling and warming curve with same equation and overlap of experimental data justifies it. This is similar to that observed by Matt et al [41] for their FeRh thin film grown on MgO substrate. We like to point out that the amplitude of the cooling curve is systematically lower (though marginal) than warming curve. Here it is worth mentioning that even though the measurements were carried out after 5 min wait after temperature stabilization,

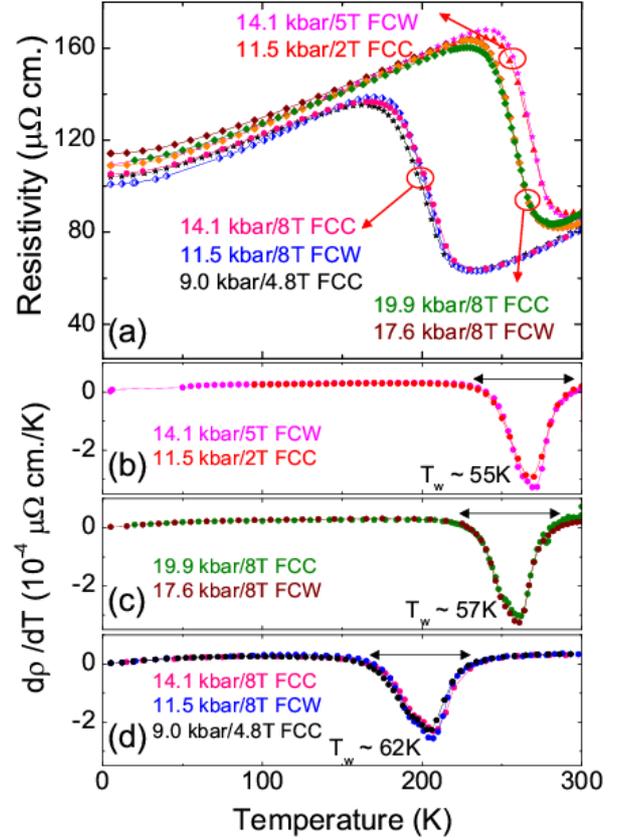

**Figure5:** *[a] $\rho$ vs. T and [b-d] $d\rho/dT$ vs. T for three set of measurements for which H and P are chosen in such a way that $T^*(H_1, P_1) = T^{**}(H_2, P_2)$. FCC and FCW indicate measurement during cooling and warming respectively after magnetic field is applied at room temperature. All the curves for which this condition is satisfied follows similar temperature dependence irrespective of cooling or warming, indicating similar nucleation and growth process during cooling and warming.*

there could be difference in the rate of temperature change during temperature sweep/stabilization between two temperatures.

Above results show that the transition width and hysteresis width are determined by the transition temperature irrespective of applied magnetic field and pressure under the condition that $T_N$ is constant for such combination of P and H. To check, if the hysteresis and transition width are solely determined by temperature we carried out isothermal magnetoresistance (MR) under various constant pressure. If temperature is the only determining factor then we expect that irrespective of applied pressure the transition width and the extent of hysteresis will be same and only effect of pressure will be on the critical field required for the AFM-FM transition. Results of some of the isothermal and isobaric MR measurements are shown in figure 6 (a-b). These curves show a giant magnetoresistance associated with magnetic field induced first order AFM-FM transition the magnitude of which increase with decreasing temperature. As expected at a constant temperature higher magnetic field is required for AFM to FM transition for higher pressure. To compare the hysteresis width and transition width at constant temperature but for different pressure, field derivative of respective curves were taken. Figure 5(b)-(d) shows the field derivative of MR curve, which are shifted along H-axis to match the peak position during field increasing cycling. The overlap between curves at constant temperature is striking and is in line with our expectation from ρ − T data. These results once again support the inference that the transition width and hysteresis width are determined by temperature. Here also the hysteresis width increases with decreasing temperature which is found to be 1.2 and 1.9 Tesla for 275 K and 175 K respectively.

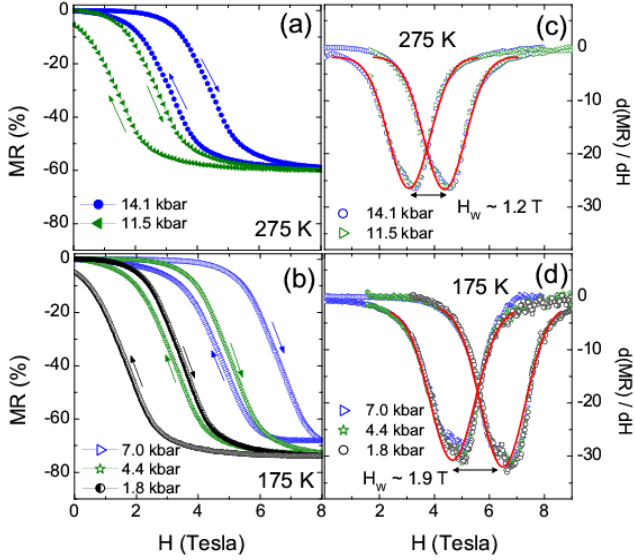

*Figure 6:* Isothermal magnetoresistance at *[a]* 175 K and *[b]* 275 K at various constant pressure along with *[c-d]* d(MR)/dH. The d(MR)/dH curves are shifted along H-axis to match the peak position. All the derivative overlap with each other showing that both hysteresis and transition width are independent of pressure. Red lines in (c-d) show Gaussian fit to d(MR)/dH curve.

It is possible that transition width may also be dependent on H and P but has not been reflected in our measurement due to limited range of these parameters. In the investigated range of these parameters the critical magnetic field required for AFM-FM transition show almost linear variation in H vs. P diagram (see figure 2(e)). It will be interesting to investigate this behavior in a region where it deviates from linearity. Finally we like to comment on the limitation of above interpretation. We have ignored the time dependence of first order transition. To avoid the relaxation effect (or time dependence) across the first order magnetic transition, we have tried to maintain the rate of temperature variation constant in all of the above measurement. We have also ignored the strain effect in interpretation of our data. In many systems supercooling and superheating curve has been shown to be governed by different nucleation and growth process due to strain effect. Such behavior has been generally observed in thin film structure [27, 41]. In such cases even supercooling and superheating curve can have entirely different shapes and will not overlap for any combination of H, P as has been observed in figure 5. It appear strain effect on studied characteristic of first order transition in our sample are minor.

### IV. CONCLUSIONS

Our study of first order AFM-FM transition in $Fe_{49}(Rh_{0.93}Pd_{0.07})_{51}$ under external pressure and magnetic field shows that the rate of change of transition temperature with pressure and magnetic field is consistent with earlier reported correlation with $T_N$. From our study this correlation appears to be valid not only for $T_N$ variation with composition but also for $T_N$ variation with magnetic field and pressure i.e. $dT_N/dP$ and $dT_N/dH$ depends on $T_N(P,H)$. The hysteresis width and the transition width are shown to be dependent on TN irrespective of applied pressure and magnetic field value. The transition width at high temperature is dominated by temperature independent part arising due to quench disorder. The temperature dependent part becomes dominant at low temperature where it increases exponentially at low temperature. The isothermal magnetoresistance curve show almost constant hysteresis and transition width irrespective of applied pressure. The dependence of hysteresis width only on temperature results in Gaussian distribution of critical field for isothermal magnetoresistance. Whereas, it result in asymmetric distribution of $T_N$ with temperature. Our results show no explicit role of pressure and magnetic field on first order transition characteristic except for shift in transition temperature. However, it will be interesting to study this system under wide range of magnetic field and pressure where critical field or pressure has non-linear variation in H-P space.

## V. ACKNOWLEDGMENTS

Sachin Kumar is acknowledged for help during measurements. Pallavi Kushwaha would like to acknowledge CSIR, India for financial support.

## VI. REFRENCES